\documentclass[galaxies,article,accept,oneauthor,pdftex,12pt,a4paper]{mdpi}

\lastpage{x}
\doinum{10.3390/------}
\pubvolume{2}
\pubyear{2015}
\history{Received: xx / Accepted: xx / \\Published: xx}
\pdfoutput=1

\usepackage{amsfonts} 
\usepackage{amstext}
\usepackage{amssymb} 
\usepackage{subfig}
\usepackage{textcomp}
\usepackage{soul}

\newcommand{\hi } {{\rm H}\,{\small\rm I} \,}

\newcommand {\apj}{Astrophys.~J.}
\newcommand {\aj}{Astrophys.~J.}

\newcommand {\apjl}{Astrophys.~J.}
\newcommand {\mnras}{Mon. Not. R. Astron. Soc.}
\newcommand {\aap}{Astron. Astrophys.}

\newcommand {\pasa}{PASA}

\newcommand {\nat}{Nature}

\Title{Tidal Dwarf Galaxies: Disc Formation at $z\simeq0$}

\Author{Federico Lelli$^{1}$, Pierre-Alain Duc, Elias Brinks \& Stacy S. McGaugh}

\address[1]{$^{1}$ Department of Astronomy, Case Western Reserve University,
10900 Euclid Ave, Cleveland, OH 44106, USA; E-Mail: federico.lelli@case.edu;
Tel.: +1-216-368-0917}

\abstract{Collisional debris around interacting and post-interacting galaxies often display condensations of gas and young stars that can potentially form gravitationally bound objects: Tidal Dwarf Galaxies (TDGs). We summarise recent results on TDGs, which are originally published in Lelli et al. (2015, A\&A). We study a sample of six TDGs around three different interacting systems, using high-resolution \hi observations from the Very Large Array. We find that the \hi emission associated to TDGs can be described by rotating disc models. These discs, however, would have undergone less than one orbit since the time of the TDG formation, raising the question of whether they are in dynamical equilibrium. Assuming that TDGs are in dynamical equilibrium, we find that the ratio of dynamical mass to baryonic mass is consistent with one, implying that TDGs are devoid of dark matter. This is in line with the results of numerical simulations where tidal forces effectively segregate dark matter in the halo from baryonic matter in the disc, which ends up forming tidal tails and TDGs.}


\begin{document}
 
\section{Introduction}\label{sec:intro}

Collisional debris around interacting and post-interacting galaxies are often observed to contain stellar and gaseous condensations that have masses, sizes, and star-formation rates (SFRs) comparable to those of dwarf galaxies \citep[][]{Weilbacher00, Boquien10}. \citet{Zwicky56} was the first to suggest that these condensations may collapse under self-gravity and form new independent entities: tidal dwarf galaxies (TDGs). Numerical simulations of interacting galaxies predict the formation of TDGs \citep{Bournaud06, Wetzstein07} and suggest that they may survive for several Gyr against internal or external disruption \citep{Ploeckinger15}. Interestingly, these simulations predict that TDGs should be free of non-baryonic dark matter (DM). This occurs because of two basic dynamical effects~\citep{Barnes92}: (i) tidal forces effectively segregate baryons in the disc (which is dynamically cold and can form tails and TDGs) from DM in the halo (which is too dynamically hot to form narrow tails); and (ii) once a TDG has formed, it has a shallow potential well and cannot accrete DM particles with typical velocity dispersions of $\sim$200 km~s$^{-1}$. Similarly, TDGs are moving at the characteristic speed of the central DM halo ($\sim$200 km~s$^{-1}$) and cannot be accreted by DM sub-haloes that have shallow potential wells ($\sim$50 km~s$^{-1}$).

\citet{Bournaud07} used high-resolution \hi observations to study the dynamics of three TDGs around NGC 5291. Puzzingly, they found that the dynamical masses ($M_{\rm dyn}$) of these TDGs are a factor of $\sim$3 higher than the baryonic masses ($M_{\rm bar}$). They hypothesised that the missing mass may consist of very cold molecular gas, which is not traced by CO observations but may constitute a large portion of baryonic DM in galaxy discs \citep[][]{Pfenniger94a}. Alternatively, it was shown that MOdified Newtonian Dynamics (MOND) can reproduce the observed rotation velocities of these TDGs without any need of DM \citep{Milgrom07, Gentile07b}. In this regard, TDGs may help to distinguish between MOND and the standard model of cosmology, given that these two paradigms predict different internal dynamics for these recycled galaxies \citep{Kroupa12}.

\begin{figure*}[t]
\centering
\includegraphics[width=0.9\textwidth]{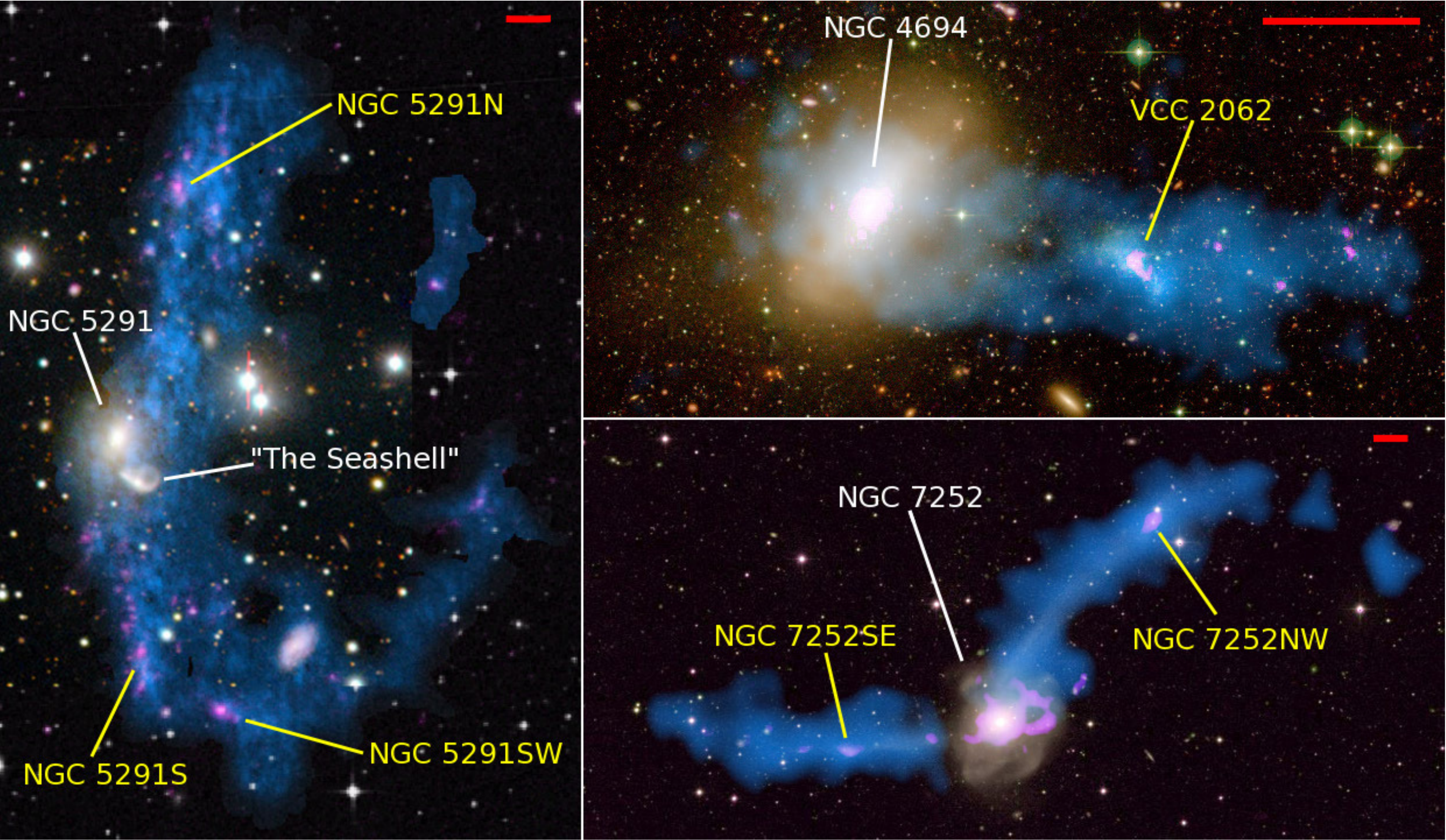}
\caption{Interacting systems with genuine TDGs (indicated in yellow). Optical images are overlaid with \hi emission from the VLA (blue) and FUV emission from GALEX (pink), tracing star-forming regions. The red bar corresponds to $\sim$10 kpc for the assumed distances.}
\label{fig:mosaic}
\end{figure*}
Here I summarise recent results on the internal dynamics of TDGs (published in \citet{Lelli15}), which are based on high-resolution \hi observations from the Very Large Array. We investigate six TDGs around three different interacting systems (Figure~\ref{fig:mosaic}): the late-stage merger NGC~7252 (``Atoms for Peace''), the collisional galaxy NGC~5291, and the post-merger lenticular NGC~4694. In particular, we revisit the DM content estimated by \citet{Bournaud07} for NGC 5291. We consider only \textit{genuine} TDGs that satisfy the following requirements: (i) they are real condensations of gas and stars, which are not due to projection effects along the debris; (ii) they have higher metallicities than galaxies of similar mass, indicating that they are not pre-existing dwarf galaxies but recycled objects forming out of pre-enriched gas \citep[][]{Duc14, Lelli15}; and (iii) they display a velocity gradient that is kinematically decoupled from the tidal debris, pointing to a local potential well \citep{Bournaud07, Lelli15}.

\section{Dynamics of TDGs}\label{sec:models}

\begin{figure*}
\centering
\includegraphics[width=0.9\textwidth]{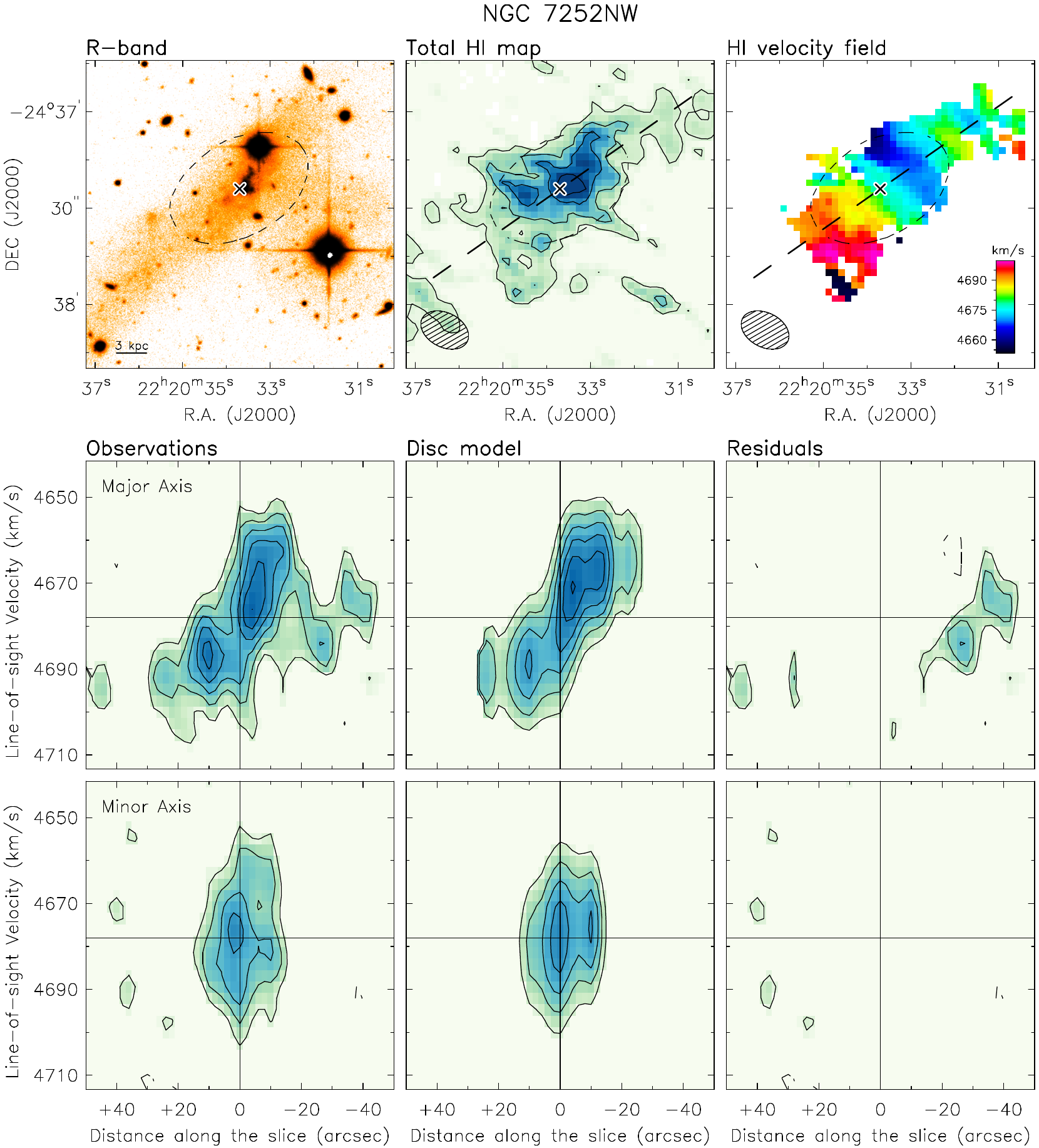}
\caption{\textbf{Top panels}: optical image (\textit{left}), total \hi map (\textit{middle}), and \hi velocity field (\textit{right}). The dashed ellipse corresponds to the disc model for the assumed inclination. The cross and dashed line illustrate the kinematical centre and major axis, respectively. In the bottom-left corner, we show the linear scale (optical image) and the \hi beam (total \hi map and velocity field). In the total \hi map, contours are at $\sim$1.5, 3.0, 4.5, and 6 M$_{\odot}$~pc$^{-2}$. \textbf{Bottom panels}: PV diagrams obtained from the observed cube (\textit{left}), model cube (\textit{middle}), and residual cube (\textit{right}) along the major and minor axes. Solid contours range from 2$\sigma$ to 8$\sigma$ in steps of $1\sigma$. Dashed contours range from $-$2$\sigma$ to $-$4$\sigma$ in steps of $-$1$\sigma$. The horizontal and vertical lines correspond to the systemic velocity and dynamical centre, respectively.}
\label{fig:N7252NW}
\end{figure*}

In \citet{Lelli15}, we study the \hi kinematics of TDGs by building 3D kinematical models. These techniques have been widely used and validated for typical dwarf galaxies \citep{Swaters09, Lelli14}. In short, we construct disc models (with given radius, rotation velocity, and velocity dispersion) and project them on the sky to obtain model cubes that can be directly compared with the observed cubes. For all TDGs, we find that rotating disc models can reproduce the observed \hi kinematics. An example (NGC~7252NW) is presented in Figure~\ref{fig:N7252NW}. The top panels show a $R$-band image (left), the \hi map (middle), and the \hi velocity field (right). Clearly, the gas distribution and kinematics resemble a rotating disc. The bottom panels show position-velocity diagrams obtained from the observed (left), model (middle), and residual (right) cubes along the major and minor axes. A disc model with axisymmetric kinematics can reproduce both the major and minor axes, suggesting that possible non-circular motions are relatively small. These techniques represent a major improvement over previous kinematic studies of TDGs because they allow us to estimate rotation velocities considering the effects of the gas distribution, velocity dispersion, disc inclination, and instrumental resolution. In particular, we find that the rotation velocities from \citep{Bournaud07} were over-estimated because the line-broadening due to turbulent motions was not taken into account.

Although a rotating disc provides a good description of the \hi kinematics, there is an important issue with this model. The orbital times at the edge of the disc are longer than the dynamical timescale for the interaction/merger event estimated from numerical simulations. Hence, these \hi discs have undergone less than a full rotation since the epoch of the TDG formation, raising the question of whether they are in dynamical equilibrium. For collisionless stellar systems, the process of relaxation takes a few orbital times due to the energy exchange between individual particles and the total gravitational potential \citep{LyndenBell67}. For dissipative gas systems as TDGs, the energy can probably be exchanged and radiated away on a local crossing time ($\sim$10 Myr), thus one may speculate that the process of relaxation in collisional systems is faster than in collisionless ones. This remains an open issue that deserves further investigation.

\begin{figure}[h!]
\begin{minipage}{0.5\textwidth}
\centering
\includegraphics[width=0.98\textwidth]{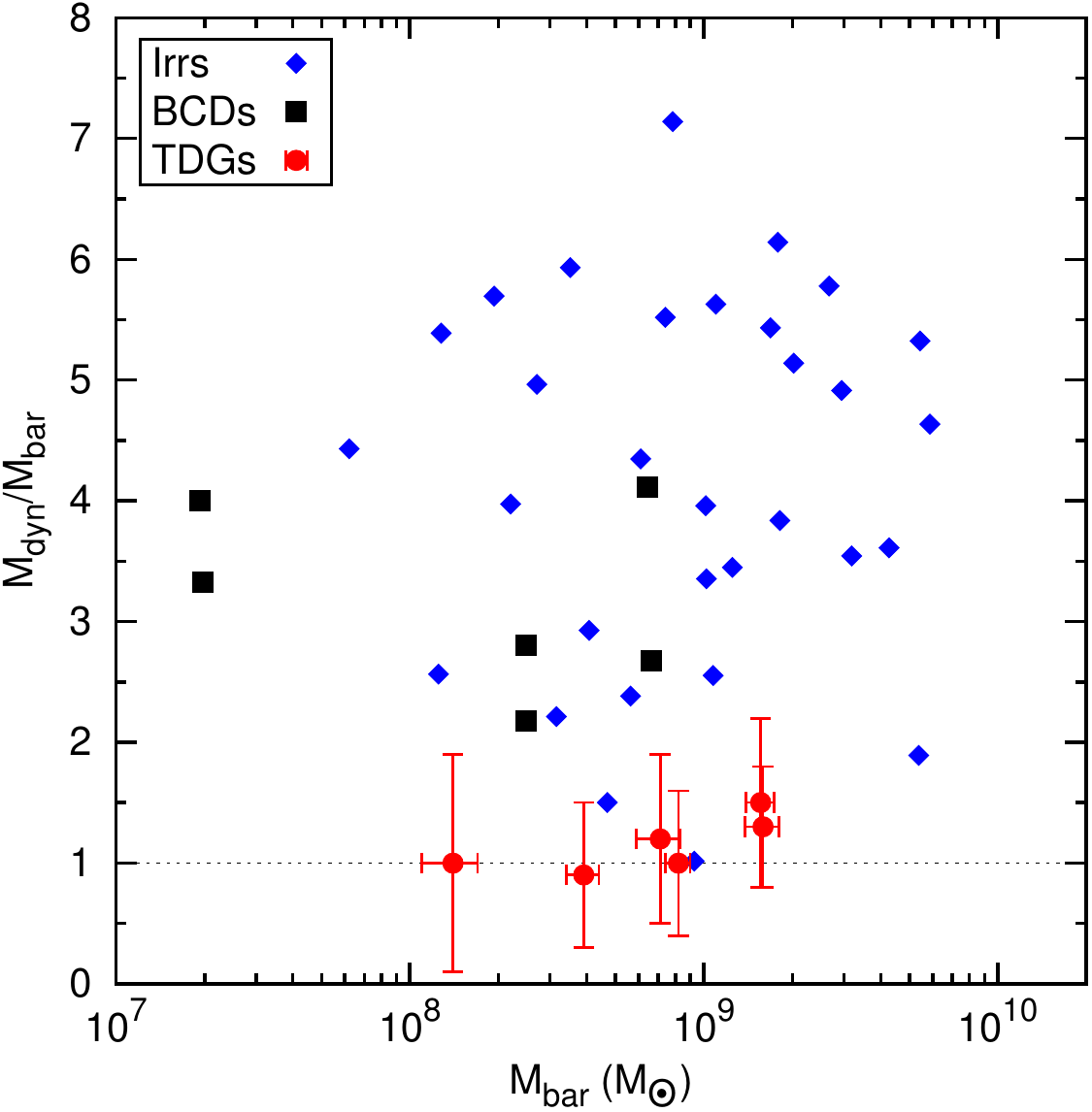}
\end{minipage}
\begin{minipage}{0.5\textwidth}
\centering
\includegraphics[width=0.87\textwidth]{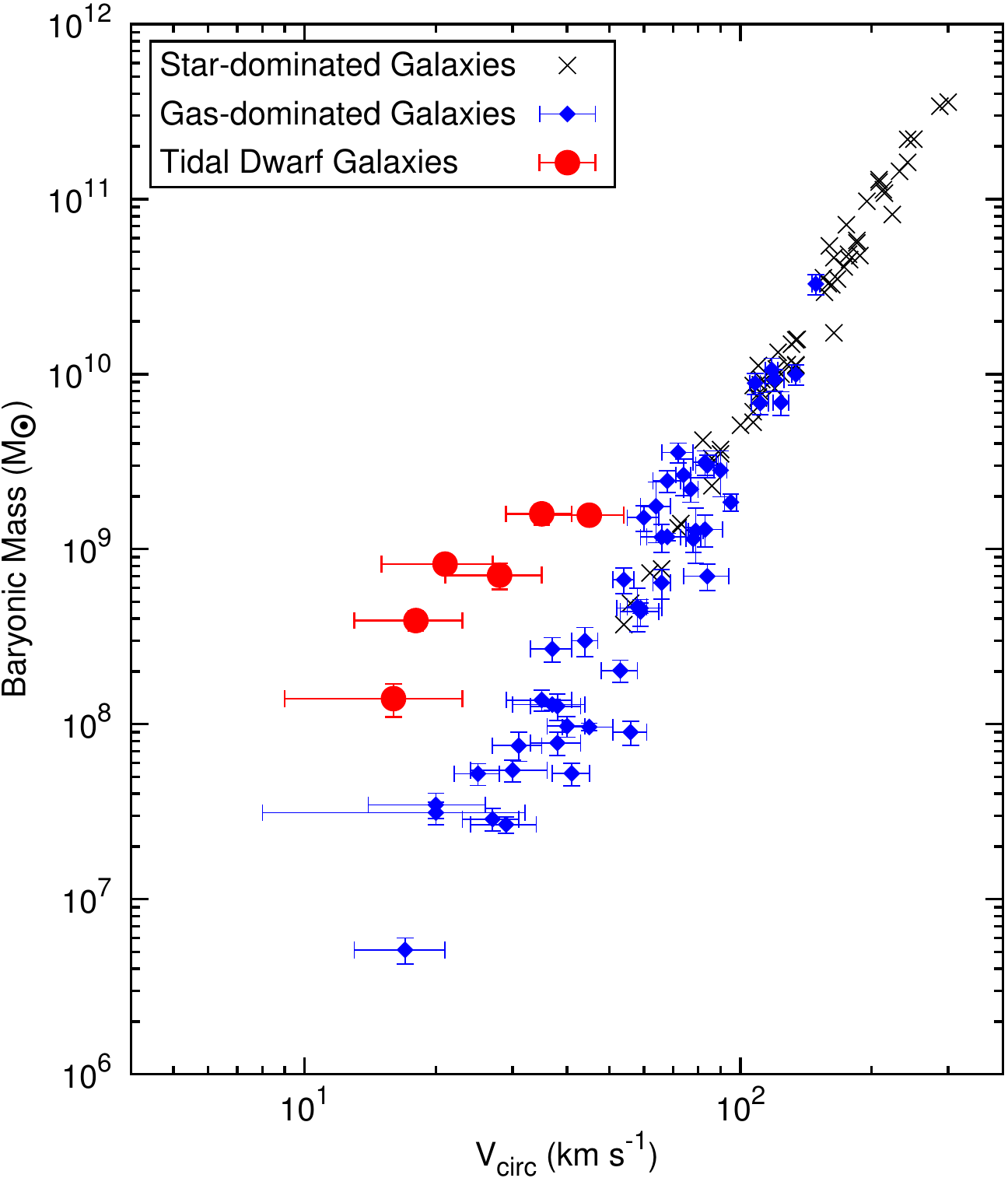}
\end{minipage}
\caption{\textbf{Left panel}: Mass budget within the optical radius for different types of dwarf galaxies (see \citep{Lelli15} for details). \textbf{Right panel}: Location of TDGs on the BTFR. $V_{\rm circ}$ is the circular velocity after correction for pressure support. For typical galaxies, $V_{\rm circ}$ is measured along the flat part of the rotation curve \citep{McGaugh05,McGaugh12}. For TDGs, the shape of the rotation curve is uncertain, thus $V_{\rm circ}$ may not correspond to the velocity along its flat part.}
\label{fig:BTFR}
\end{figure}
If we assume that TDGs are in equilibrium, we can estimate dynamical masses using the observed rotation velocities (after correction for pressure support). We also estimate baryonic masses as the sum of atomic gas (1.33 $M_{\hi}$), molecular gas ($M_{\rm mol}$), and stars ($M_{*}$). In TDGs $M_{\rm bar}$ is heavily dominated by atomic gas, hence typical uncertainties on $M_{\rm mol}$ (due to the CO-to-H$_2$ conversion factor) and $M_{\rm *}$ (due to the stellar mass-to-light ratio) do not affect our results. We find that the inferred dynamical masses are consistent with the baryonic masses, suggesting that TDGs are nearly devoid of DM. This is illustrated in Figure~\ref{fig:BTFR} (left), where we compare the $M_{\rm dyn}/M_{\rm bar}$ ratio of TDGs with that of dwarf irregulars \citep[Irrs; blue dots from ][]{Swaters09} and starburst dwarfs \citep[BCDs; black squares from ][]{Lelli14}. TDGs have smaller values of $M_{\rm dyn}/M_{\rm bar}$ than typical dwarfs, pointing to an apparent lack of DM. In particular, we do not confirm the high values of $M_{\rm dyn}/M_{\rm bar}$ from \citep{Bournaud07} for the TDGs around NGC~5291. This happens because we estimate (i) lower values of $V_{\rm rot}$ due the technical reasons discussed above, and (ii) higher values of $M_{\rm gas}$ since we integrate the \hi flux out to larger radii than \citep{Bournaud07} (as suggested by our 3D disc models).

\section{Implications for the dark matter distribution}

If TDGs are in dynamical equilibrium, their low values of $M_{\rm dyn}/M_{\rm bar}$ provide strong support to numerical simulations where tidal forces segregate halo material from disc material. They also provide an indirect but empirical confirmation to the hypothesis that DM is distributed in pressure-supported haloes. The internal dynamics of TDGs also constrains putative ``dark discs'' (either baryonic or non-baryonic) in the progenitor spiral galaxies, given that any kind of rotation-supported material may form tidal tails and be accreted in TDGs. For example, it has been argued that spiral galaxies should possess a rotating disc of non-baryonic DM, which is formed by the near-plane accretion of DM-dominated satellite galaxies \citep{Read08}. Our results imply that either these dark discs are significantly less massive than the baryonic one or they are too dynamically hot to be accreted by TDGs.

\citet{Pfenniger94a} proposed very cold molecular gas as an alternative to non-baryonic DM in disc galaxies. This putative dark gas would consist of small ``clumpuscules'' that do not emit any detectable line and are not associated with star formation. The existence of these cold H$_2$ clumpuscules has been advocated to explain the origin of spiral patterns and long-lived warps in the outer regions of \hi discs \citep{Revaz04, Revaz09}. Our current results do not support the existence of large numbers of cold H$_{2}$ clumpuscules. Similarly, other forms of baryonic DM in the progenitor disc galaxies (like brown dwarfs, low-mass stars, or stellar remnants) seem very unlikely. Our errors on $M_{\rm dyn}/M_{\rm bar}$, however, are consistent with the expectations for ``standard'' dark gas, such as optically thick \hi or photodissociated CO in the outer regions of molecular clouds. For the Milky Way, the Planck collaboration has estimated that the mass of dark gas may be comparable to that of CO-traced molecular gas \citep{Planck11a}. In principle, the values of $M_{\rm dyn}/M_{\rm bar}$ in TDGs may be used to identify the existence of dark gas in the progenitor galaxies, but the current uncertainties on $M_{\rm dyn}$ do not allow such a test.

\section{The baryonic Tully-Fisher relation and MOND}

TDGs seem to systematically deviate from the BTFR (Fig.~\ref{fig:BTFR} right). They show small rotation velocities for their baryonic masses, in line with the observed lack of a DM halo. We stress, however, that the data-points from \citep{McGaugh05, McGaugh12} consider the asymptotic velocity along the flat part of the rotation curve ($V_{\rm flat}$), which minimises the scatter on the BTFR. For the TDGs in our sample, the shape of the rotation curve is uncertain, thus it remains unclear whether we are tracing $V_{\rm flat}$.

The universality of the BTFR is a key prediction of MOND \citep{Milgrom83}. According to MOND, disc galaxies can deviate from the BTFR only if (i) they are out of equilibrium \citep{McGaugh10} or (ii) they are affected by the external-field effect \citep[EFE,][]{Bekenstein84}. Both circumstances may occur here. In \citet{Lelli15}, we perform a first-order analysis of the EFE, considering the gravitational field of the progenitor galaxies. We find that the EFE can substantially decrease the velocities expected in MOND with respect to the isolated case (corresponding to the BTFR), but a systematic discrepancy persists between MOND and observations (at the level of $\sim$1-2$\sigma$ for individual galaxies). We stress that our EFE analysis formally applies to the one-dimentional case and neglects the direction and time-variation of the external field. Numerical simulations are needed to investigate the EFE in detail and address the process of TDG formation and relaxation in the context of MOND.

\acknowledgements{Acknowledgments}

The work of FL and SSM is made possible through the support of a grant from the John Templeton Foundation.






\end{document}